\documentclass[aps,prl,twocolumn,groupedaddress,showpacs]{revtex4}

\usepackage{graphicx}% Include figure files

\begin{document}

\title{Global Analysis of Synchronization in Coupled Maps}

\author{J\"urgen Jost and Kiran M. Kolwankar}
\affiliation{Max Planck Institute for Mathematics in the Sciences,
Inselstrasse 22, 
D-04103 Leipzig, Germany}

\date{\today}

\begin{abstract}
We introduce a new method for determining the global stability of
synchronization in systems of coupled identical maps. The method is
based on the study of invariant measures. Besides the
simplest non-trivial example, namely two symmetrically coupled tent
maps,  we also treat the case of two asymmetrically coupled  tent maps
as well as a globally coupled network.
Our main result is the identification of the precise value of the coupling parameter
where the synchronizing and desynchronizing transitions take place.
\end{abstract}


\maketitle

\section{Introduction}
Although   the phenomenon
of synchronization of nonlinear oscillators was discovered
already in 1665 by Huyghens, it has received systematic attention only
recently~\cite{PRK,Pecora1997},  
owing mainly
to newly found applications and the understanding of nonlinear 
systems that we have achieved through modern methods.
The field received an impetus after it was discovered by Fujisaka-Yamada~\cite{Fujisaka1983} and Pikovsky~\cite{Pikovsky1984} and further analyzed by Pecora-Carroll~\cite{Pecora1990} 
that even  chaotic trajectories can synchronize. Since then
the phenomenon has found applications in secure communication,
coupled Josephson junctions arrays etc. 
From a completely different perspective, it was observed~\cite{Eckhorn1988,Gray1989} 
experimentally that synchronous firing of neurons was important in
feature binding in neural networks. 
Distant groups of neurons rapidly synchronize and desynchronize 
as the input stimulus changes, each synchronized group representing a
collection of features belonging to the same percept.
Thus a systematic understanding 
of synchronization and desynchronization
has become important in neural information 
processing~\cite{Hansel1992,Buzsaki2004,Friedrich2004}.   

As a result of these developments various coupled dynamical systems
have been studied that are either continuous or discrete in time, 
either continuously
coupled or pulse coupled etc.
and different types of synchronizations have arisen such as phase
synchronization or complete synchronization, etc.
Coupled maps~\cite{Kaneko1993} have been used to a great extent to understand
synchronization in particular~\cite{Gade1996,jost:016201,jalan:014101,
Celka1996,Chatterjee1996,Maistrenko1996,Balmforth1999,
Gade1995,Liu1999,Manrubia1999,
Sinha2002}
and complexity in dynamical systems~\cite{Gaspard1993,Olbrich2000,monte:254101}
in general. The paradigm here consists of identical individual
maps, typically iterates of a functional equation like the logistic
one that produce chaotic dynamics. These maps then are coupled, that
is each of them computes its next state not only on the basis of its
own present state but also on the basis of those other ones that it is
coupled to. With appropriate conditions on the coupling strengths, the
individual solution is also a solution of the collective dynamics,
that is, when all the maps follow their own intrinsic dynamics
synchronously, we also have a solution of the coupled dynamics. This
is the simplest case of synchronization. That
synchronized collective dynamics, however, need not be stable even if
the individual ones are. Therefore, the study of 
 synchronization essentially depends on a
stability analysis. 
One may use linear stability analysis
or global stability analysis (for some recent formulations, 
see~\cite{Pecora1998,jost:016201,chen:026209}). 
In the linear stability analysis one
assumes that the synchronized state follows the same dynamics as the
individual uncoupled dynamics and then studies the stability of the
dynamics transverse to the synchronizing manifold. 
However it has been observed~\cite{Atay2004} in coupled systems
with delays that the synchronized dynamics can be quite different
from the individual uncoupled dynamics.
Also, it has been demonstrated~\cite{Ginelli2002,Ginelli2003}
that the linear stability analysis
can actually fail in a nonlinear setting, due to global nonlinear effects.
For a global stability analysis, 
one has to guess a Lyapunov function which then gives a criterion for
synchronization. Thus the linear stability takes only the local dynamics
into account and it makes a restricting assumption on the synchronized
dynamics. In contrast, the global stability analysis, while global, depends
on the choice of the Lyapunov function and may not always give 
optimal criteria. 
In general optimal global results are difficult to obtain. In many
cases numerical results indicate the global stability of the synchronized
state, but it is difficult to prove stability rigorously.
Thus it is necessary to develop a method to study
synchronization which goes beyond these drawbacks. That is, it should
be global and should not make any reference to the synchronized dynamics.   
This is what we achieve in this paper. We present a new global
stability method that leads to sharp results, that is, it can
determine the critical value of the coupling parameter above which
global synchronization occurs. In order to exhibit the principal
features, we first apply the method to the simplest non-trivial case,
namely two symmetrically coupled tent maps. 
Then we consider asymmetric coupling and a globally coupled
network.

%In subsequent, more
%detailed work, we shall treat more general cases, like general
%coupling schemes, unsymmetric coupling, and, in particular, also other
%maps, like the logistic map, the prototypical example for the generation
%of chaotic behaviour. 

Our method uses invariant measures or stationary
densities\cite{LM} to study synchronization.
A stationary density is the density obtained
by starting from some initial distribution of the initial conditions
and letting the system evolve for a long time. Though the complete characterization
of such densities can be difficult,  we shall demonstrate below that  
it may suffice to study
its support to understand synchronization and this can be easier than  studying the whole measure. Clearly, this
approach is inherently global.  The key idea  is to study the support 
of the invariant measure for different coupling strengths 
and to determine the critical value of the coupling strength
when the support shrinks to the synchronized manifold.
 We consider the
following coupled dynamical system
\begin{eqnarray}
X_{n+1}=Af(X_n) =: S(X_n)
\end{eqnarray}
where $X=(x_1,x_2,\cdots , x_N)^T$ is an $N$-dim. column vector, 
$A$ is an $N\times N$ 
coupling matrix and $f$ is a map from $\Omega=[0,1]^N$ onto
itself. As already mentioned, in the present paper we consider the tent map defined as:
\begin{eqnarray}
f(x) = \left\{ \begin{array}{ll}
2x & 0\leq x \leq 1/2 \\
2-2x & 1/2 \leq x \leq 1.
\end{array} \right.
\end{eqnarray}
Different coupling matrices are chosen. 

\section{Invariant measures}{\label{se:im}}
A measure $\mu$ is said to be invariant under a transformation
$S$ if $\mu(S^{-1}(A)) = \mu(A)$ for any measurable subset $A$ of $\Omega$.
For example, a Dirac measure concentrated at a
fixed point of the transformation is clearly invariant. An invariant
measure need not be unique as can be seen immediately from the fact
that if there are several fixed points then the  different Dirac
measures at these  fixed points as well as their linear combinations
are all invariant. A natural way to 
obtain invariant measures is to simply iterate any measure under the
transformation $S$ and take an asymptotic (weak-$\star$) limit of
these iterates. A sequence $\mu_n$ of measures converges in the
weak-$\star$ sense to a measure $\mu$ when for all continuous
functions $f$ on $\Omega$
\begin{equation}
\lim_{n \to \infty} \int_\Omega f(x)\, \mu_n(dx)=\int_\Omega f(x)\,
\mu(dx).
\end{equation}
The usefulness of this concept derives from the fact that bounded sets
are weak-$\star$ precompact, that is, contain a weak-$\star$
convergent subsequence, see e.g. \cite{Rudin}. In particular, this applies to the dynamical
iterates of some initial measure. 
For example, one may start with any Dirac measure. 
Our
interest, however, is not in such singular measures as they do not
sample the whole phase space. We would like to start from a
distribution of initial conditions spread over the whole phase
space (say uniformly) and study its evolution and what kind of
asymptotic limit it leads to. This is the idea of the SRB-measures, as
they are called after Sinai, Ruelle, and Bowen. 
For some dynamical systems, any invariant measure is 
singular. In such cases even if we start with a uniform
density we obtain a singular measure asymptotically. 
However, if, for example,  the
map is expanding everywhere then an SRB measure is absolutely
continuous w.r.t. to Lebesgue measure. Therefore
here we restrict ourselves to one particular map that is expanding, namely
the tent map.
We want to investigate how such a density depends on  the coupling
strength and 
would like to see when  its support shrinks to the 
synchronization manifold.

There are various ways to find  invariant measures. One approach
is to use the so called Frobenius-Perron operator. It is defined as: 
\begin{eqnarray}
\int\int_D P\rho(X') d^NX' = \int\int_{S^{-1}(D)} 
\rho(X') dX'
\end{eqnarray}
If we choose $D=[0,x_1]\times\cdots \times[0,x_N]$ then we get
\begin{eqnarray}
P\rho(X)  =\left( {\partial\over\partial x_1} \cdots
{\partial\over\partial x_N}\right) \int\int_{S^{-1}(D)} 
\rho(X') dX'
\end{eqnarray}
Our $S$ is not invertible. In fact, it has $2^N$ disjoint parts.
Let us denote them by $S_i^{-1}$, $i=1,...,2^N$. If $X\in\Omega$,
since $f$ is symmetric, we get 
\begin{eqnarray}
P\rho(X) = J^{-1}(X) \sum_{i=1}^{2^N} \rho(S_i^{-1}(X))
\end{eqnarray}
where $J^{-1}(X)=|dS^{-1}(X)/dX|$. 

\section{Symmetrically coupled maps}{\label{se:scm}}

In this section we choose a symmetric dissipative coupling 
given by
\begin{eqnarray}
A=\left( \begin{array}{cc}
a-\epsilon & \epsilon \\
\epsilon & a-\epsilon
\end{array} \right)
\end{eqnarray}
where $0< \epsilon<1$ is the coupling strength and $0<a<1$ is a
parameter. This choice for $A$ also satisfies the constraint that
the row sum is a constant. This guarantees that the synchronized
solution exists.

\subsection{The case $a=1$}
With this choice of $A$ we get the following functional equation for 
the density.
\begin{eqnarray}
P\rho(x,y)\!\!\! &=&\!\!\! {1\over{4|1-2\epsilon |}}[
\rho(\beta x/2-\gamma y/2, -\gamma x/2+\beta y/2)
\nonumber \\
&&\!\!\!+\rho(1-\beta x/2+\gamma y/2, -\gamma x/2+\beta y/2)
\nonumber \\
&&\!\!\!+\rho(\beta x/2-\gamma y/2,1+\gamma x/2-\beta y/2)
\nonumber \\
&&\!\!\!+\rho(1-\beta x/2+\gamma y/2,1+\gamma x/2-\beta y/2)
]
\end{eqnarray}
where $\gamma=\epsilon/1-2\epsilon$ and $\beta = 1+\gamma$.
Since we know that a point belonging to $\Omega$ does not
leave $\Omega$, all the arguments of $\rho$ on the right hand
side of the above equation should be between 0 and 1. This gives
us four lines 
$0\leq \beta x/2-\gamma y/2 \leq 1$ and 
$0\leq -\gamma x/2+\beta y/2 \leq 1$
which bound an area, say $\Gamma$. 
The support of the invariant measure should be contained in $\Gamma \cap \Omega$.

\begin{figure}
\includegraphics[scale=0.5]{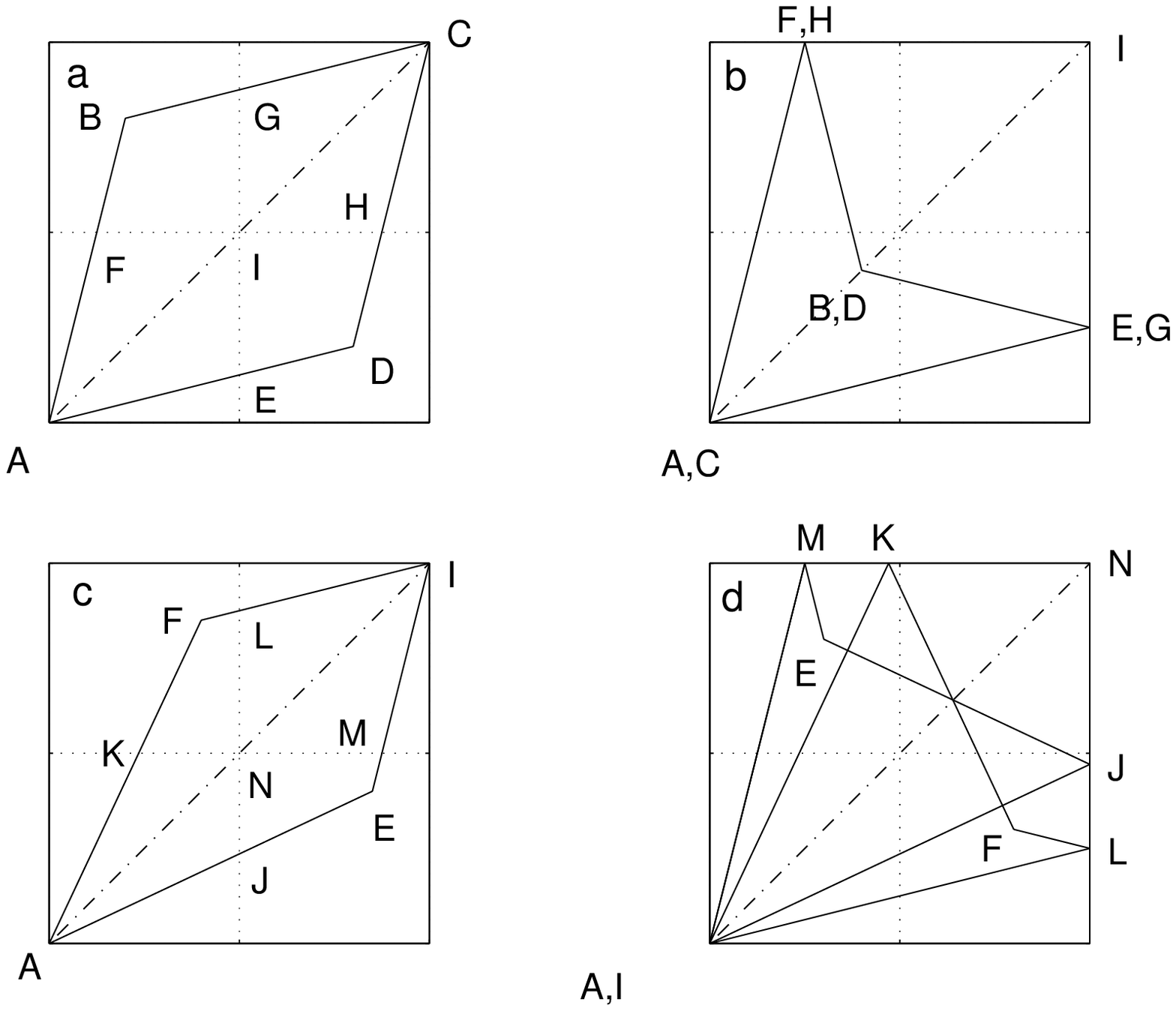}%
\caption{\label{fig1} The evolution of the support of the invariant 
measure obtained by starting from the Lebesgue measure supported on
the square $[0,1]\times [0,1]$ with $\epsilon < 1/4$. 
a) The parallelogram ABCD obtained after the first
application of $A$ b) The rhombus AFIE obtained by applying $f$ to the
rectangle ABCD in a. c) The rhombus AFIE one gets when $A$ is applied to
the rhombus in b. d) The next application of $f$ leads to the rhombus
AMNL.}
\end{figure}
In order to study the evolution of  supp $\rho$ we study the iterates
of $\Omega$ under $S=Af$. $f$ maps $\Omega$ onto $\Omega$. Application
of $A$ then leads to the parallelogram shown in Fig.~\ref{fig1}a. The next 
application of $f$ leads to Fig.~\ref{fig1}b. We use the following convention
for labelling 
the points throughout. A point is labelled by the same letter in its
image. As a result, some points in Fig.~\ref{fig1}b have two labels. 
We note that the boundary in Fig.~\ref{fig1}b is completely determined by
the part of the area in the upper right quadrant in Fig.~\ref{fig1}a.
This is essentially because $l(GI)>l(EI)$ and $l(HI)>l(FI)$.
Let us call this {\em condition A}. 
So the areas in other quadrants get mapped inside the image
of this area. Now we 
operate $A$ again to obtain Fig.~\ref{fig1}c 
and $f$ again to get Fig.~\ref{fig1}d. 
Here we have assumed that $\epsilon$ is sufficiently small so that
the $x$-coordinate of F in Fig.~\ref{fig1}c 
is less than 1/2 and by symmetry the
$y$-coordinate of E
is less than 1/2 ({\em condition B}). 
One can notice that, although the internal structure of the density
becomes more complex, the boundary is the same as that in Fig.~\ref{fig1}b.
Also, by continuing this procedure one can see that the quadrilateral AMNL
in  Fig.~\ref{fig1}d is the support of the density which remains invariant
provided $\epsilon$ is sufficiently small\footnote{It is not important
whether we call the quadrilateral $AMNL$ in Fig.~\ref{fig1}d 
an invariant area or the quadrilateral
$AFIE$ in Fig.~\ref{fig1}c since $S^n$ can also be written as $AS'^{n-1}f$
where $S'=fA$}.

The coordinates of F are given by
\begin{eqnarray}
\left(\begin{array}{cc}
1-\epsilon & \epsilon \\
\epsilon & 1-\epsilon
\end{array}\right)
\left(\begin{array}{c}
\epsilon \over{1-\epsilon} \\
1
\end{array}\right)
=
\left(\begin{array}{c}
2\epsilon  \\
1-2\epsilon+2\epsilon^2\over{1-\epsilon}
\end{array}\right)
\end{eqnarray}
As a result, the condition B leads to
$\epsilon \leq 1/4$. So when $\epsilon$ satisfies this 
condition we get the
above area as an invariant area.
In~\cite{Pikovsky1991}, Pikovsky and Grassberger have made a 
similar observation for an asymmetric generalisation of the
tent map and later Glendenning~\cite{Glendinning2001} 
has carried forward the analysis mathematically. 
However there is an important difference between the work 
in~\cite{Glendinning2001} and our work.
The aim of the ref.~\cite{Glendinning2001} is to 
study the blowout bifurcation and the Milnor attractor. As a result,
there the analysis is restricted to the range of coupling parameters
between the blowout bifurcation transition and the complete
synchronization transition (this range is absent in the case
of the symmetric tent map), whereas we are studying the transition to complete synchronization.
We carry out the complete
analysis using purely geometric arguments preserving the 
global nature of the result. In the following two sections we shall 
also extend the analysis to more general situations.

We remark that the procedure carried out so far is applicable to a
general class of nonlinear maps. Thus, by this method, one immediately obtains
a value of $\epsilon$ below which there is no synchronization
provided one can argue that there exists an absolutely continuous
invariant measure. Further discussion of this point is postponed
to the concluding section.

\begin{figure}
\includegraphics[scale=0.5]{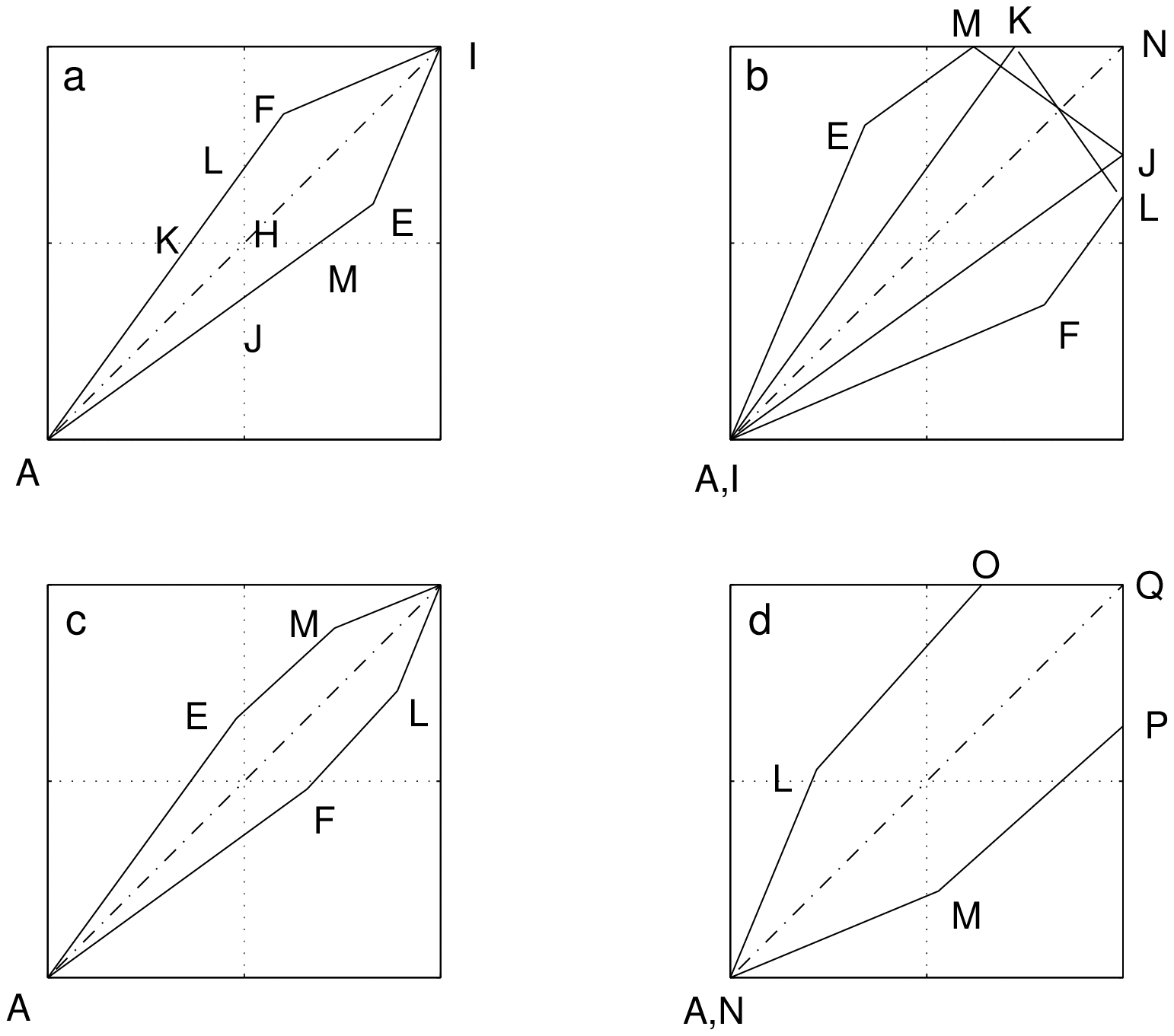}%
\caption{\label{fig2} The evolution of the support of the invariant 
measure obtained by starting from the Lebesgue measure supported on
the square $[0,1]\times [0,1]$ with $\epsilon > 1/4$. 
a) The rhombus AFIE obtained by applying $A$ to the rhombus AFIE
in Fig.~\ref{fig1}b. b) The polygon AEMNLF one gets after the application
of $f$ to the rhombus AFIE of a. c) The next application of $A$
transforms the polygon of b to this polygon AEMNLF. d) The new
polygon ALOQPM obtained by the next application of $f$ to the
polygon in c.}
\end{figure}
Now we have to check what happens for $\epsilon > 1/4$. So we
carry out the iterations with this condition in mind. As a
result in place of Fig.~\ref{fig1}c we obtain 
Fig.~\ref{fig2}a. Now if we apply $f$ we get Fig.~\ref{fig2}b. 
We see that the boundary
AEMNLF we obtain now is slightly different from that in Fig.~\ref{fig1}d, the 
corners having been chipped off.
The next application of $A$ gives us Fig.~\ref{fig2}c. It can be easily checked
explicitly that the point E lies on the left and the point M lies
on the right of the line $x=1/2$ for any $\epsilon$ greater than 
$1/4$~\footnote{This observation is true only for the map chosen, and
hence the following analysis does not directly apply to other maps.}.
This implies that the boundary obtained by the next application of $f$ 
(Fig.~\ref{fig2}d)
looks similar, that is, the number of vertices remains the same.
In fact, except for the slanted portions it is exactly the same and
we set out to argue that it is indeed different on those small
portions. That is we cannot have a set with such a boundary as
an invariant area.

In order to do this, let us first note that the 
magnitude of the slope of any
line of the form $y=mx+c$ does not change with $f$ as it scales
both axes by the same factor. And $A$ maps such a line to a new
line given by 
\begin{eqnarray}
y={{\epsilon + (1-\epsilon)m}\over{1-\epsilon+m\epsilon}}x
-{{\epsilon c({\epsilon + (1-\epsilon)m)}\over{1-\epsilon+m\epsilon}}}
+(1-\epsilon)c
\end{eqnarray}
Now if the slanted portion is stable then its slope shouldn't change
under $A$, so we have
\begin{eqnarray}
{{\epsilon + (1-\epsilon)m}\over{1-\epsilon+m\epsilon}}=m
\end{eqnarray}
This implies $m^2=1$ ($m=1$ in our case since we are interested in
the region where both coordinates are greater than $1/2$ and $f$
does not change the slope in this region). So the slanted portion
in the Fig.~\ref{fig2}c, if it is stable, has to be of the form 
$y=x-(1-2\epsilon)c$. $f$ changes this intercept to $-2(1-2\epsilon)c$
and for it to be stable this should be equal to $-c$ owing to the
symmetry. This implies that either $b=0$ or $\epsilon=1/4$.
Therefore such a slanted portion cannot be stable for $\epsilon >1/4$
and the whole area cannot be stable either. 
This shows that $\epsilon_c=1/4$ is the critical value below which
there exists an invariant area and hence no synchronization.
When $\epsilon$ crosses this $\epsilon_c$ the area becomes unstable
and collapses the synchronization manifold to the line $x=y$ and
we get synchronization.

\begin{figure}
\includegraphics[scale=0.4]{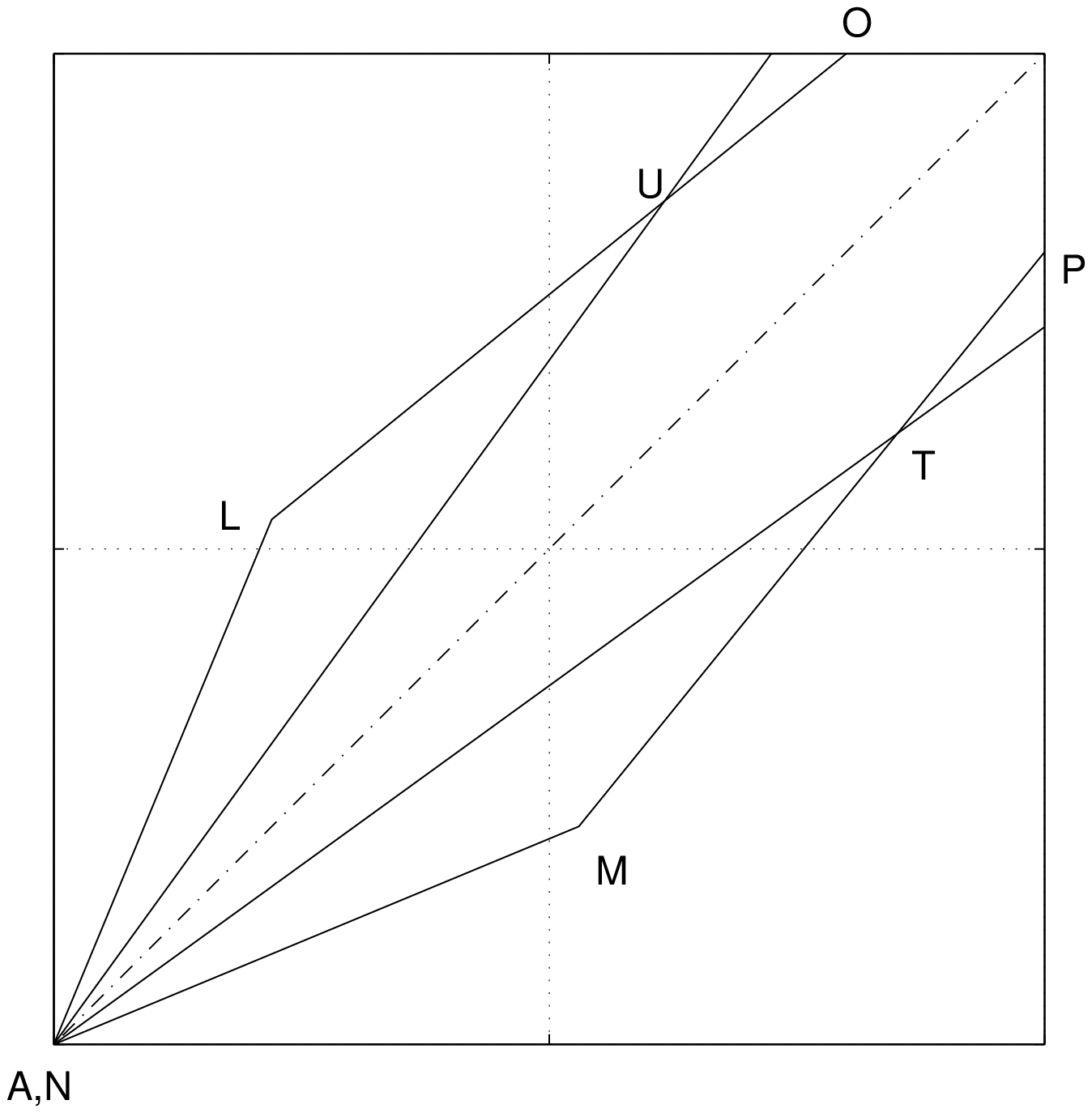}%
\caption{\label{fig3} This polygon would be obtained as the support
of the evolving measure during the course of iterations if the 
condition  A (see the text) was violated.}
\end{figure}
If the condition A  is
violated as well in Fig.~\ref{fig2}c during further iterations
then in that case the boundary
would look something like the one shown in Fig.~\ref{fig3} 
instead of that in Fig.~\ref{fig2}d.
But since $\angle MNL > \angle EAF$, there is always a part of
nonzero length of the segment $MP$ in Fig.~\ref{fig2}c which appears as a 
boundary in Fig.~\ref{fig3} (segment MT). And then the above argument will
apply to this segment.

By symmetry it can be argued that this area again becomes stable if
we increase $\epsilon$ further at $\epsilon=3/4$ and this leads to
desynchronization. Needless to say that the result agrees with that
obtained by the linear stability analysis.

It is interesting to note that the transitions
of synchronization and desynchronization are discontinuous in the
sense that the area of the support
abruptly changes  value at the critical thresholds.

Finally, we end this section by making a couple of observations
which might be useful in future refinements and/or generalizations
of this analysis to other maps. Firstly, it is interesting to
note that the vertex $F$ in Fig~\ref{fig1}b or c becomes a periodic
point of period 2 exactly at $\epsilon=1/4$ and it looses its
stability for larger $\epsilon$. Secondly, for $\epsilon < 1/4$,
the boundary of the invariant area is generated from the boundary
of $\Omega$ and hence is stable whereas for $\epsilon > 1/4$ there
is a small portion of the boundary (segment $EM$ in the 
Fig.~\ref{fig2} b or c) which does not satisfy this property
and becomes unstable.

\subsection{The case of general $a$}
In this case the synchronized dynamics is different from that
of the individual maps. This can be seen by taking the initial
point as $X_0=(z,z)$ which lies 
on the synchronization manifold which then leads to
the dynamics governed by $X_{n+1}=af(X_n)$.

\begin{figure}
\includegraphics[scale=0.5]{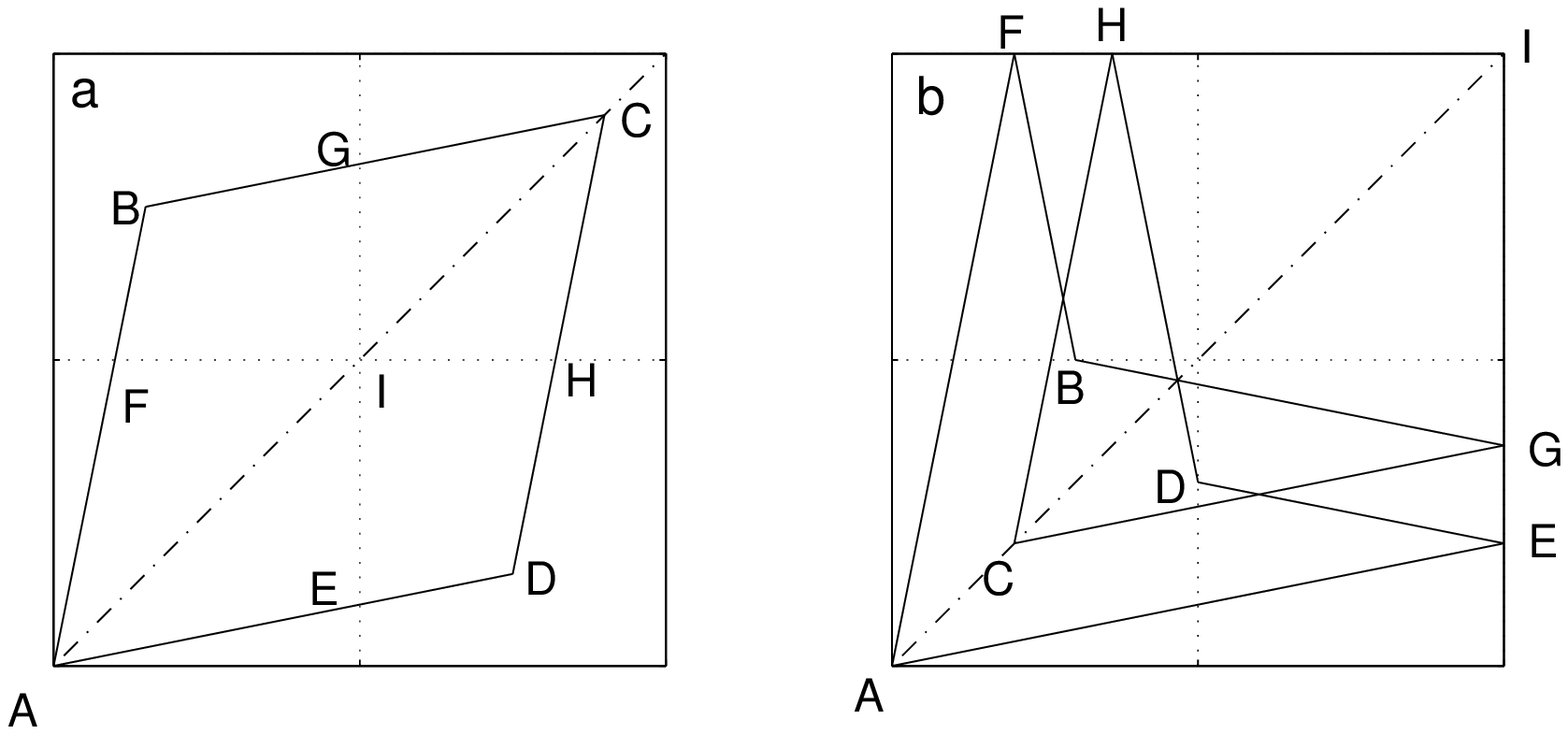}%
\caption{\label{fig4} The evolution of the support of the measure
for the coupling matrix
$A$ with general $a$. The iterates are started from the square 
$[0,1]\times [0,1]$ and $\epsilon < (2a-1)/4$. 
a) The parallelogram ABCD obtained after the first
application of $A$. b) The next application of $f$ transforms the
parallelogram in a to this figure.}
\end{figure}
\begin{figure}
\includegraphics[scale=0.4]{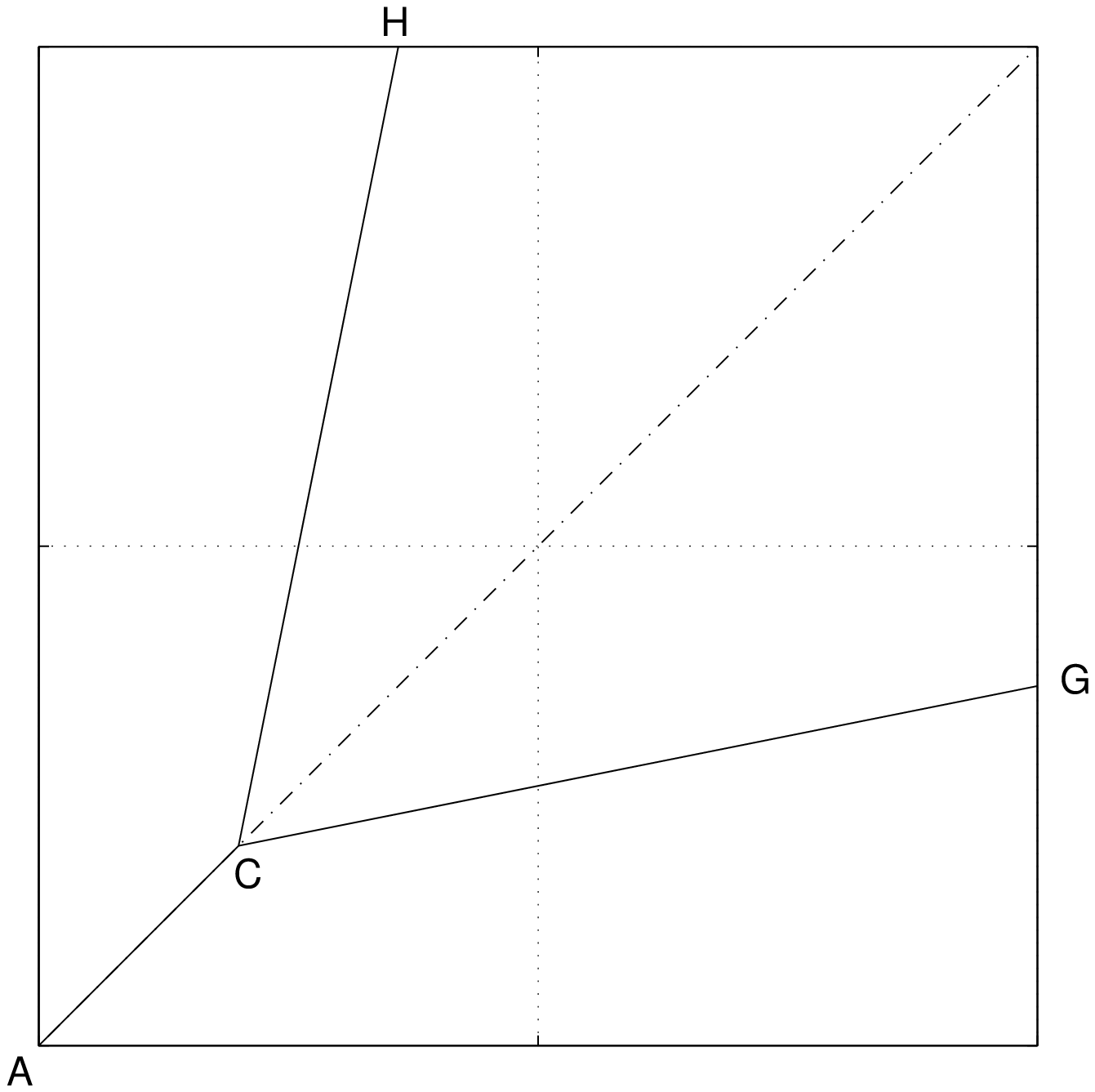}%
\caption{\label{fig5} The invariant area CHIG obtained asymptotically
for the evolution with the coupling matrix $A$ with general $a$.}
\end{figure}

Now we start the iterations of $\Omega$. The first application
of $A$ leads to the Fig.~\ref{fig4}a and then $f$ maps it to Fig.~\ref{fig4}b.
One can see that if $\epsilon$ is sufficiently small so that
the point $H$ remains on the left of the line $x=1/2$ after
application of $A$ then the $\angle HCG$ remains the same in
the subsequent iterations whereas the $\angle FAE$ decreases
with the iterations. This leads to the Fig.~\ref{fig5} asymptotically.

The above condition on $\epsilon$ can be calculated as for the
case $a=1$ and we get $\epsilon \leq (2a-1)/4$. By following 
similar arguments as for the $a=1$ case it can be shown that
this area is unstable for $\epsilon > (2a-1)/4$ and it reduces
to the line $x=y$. Again by symmetry it can be seen that the
desynchronizing transition will take place at $\epsilon = (2a+1)/4$.  

\section{Asymmetrically coupled maps}
In this section we again choose $N=2$ but now
\begin{eqnarray}
A=\left( \begin{array}{cc}
1-\epsilon_1 & \epsilon_1 \\
\epsilon_2 & 1-\epsilon_2
\end{array} \right)
\end{eqnarray}
where $0 < \epsilon_1, \epsilon_2 < 1$. We write 
$S(\epsilon_1, \epsilon_2)=Af$. We can immediately see that
\begin{eqnarray}
S(\epsilon_1, \epsilon_2)
\left(\begin{array}{c}
x \\
y
\end{array}\right)
=R S(\epsilon_2, \epsilon_1)
\left(\begin{array}{c}
y \\
x
\end{array}\right),
\end{eqnarray} 
where
\begin{eqnarray}
R=\left( \begin{array}{cc}
0& 1\\
1& 0
\end{array} \right),
\nonumber
\end{eqnarray}
i.e., the system remains unchanged under the transformation
$\epsilon_1 \leftrightarrow \epsilon_2$ and $x \leftrightarrow y$.
This implies that, if $\rho_{\epsilon_1, \epsilon_2}(x,y)$ and
$\rho_{\epsilon_2, \epsilon_1}(x,y)$ are the stationary 
densities generated by $S(\epsilon_1, \epsilon_2)$ and
$S(\epsilon_2, \epsilon_1)$ respectively,
\begin{eqnarray}
\rho_{\epsilon_1, \epsilon_2}(x,y)
=\rho_{\epsilon_2, \epsilon_1}(y,x).
\end{eqnarray}
As a result, $S(\epsilon_1, \epsilon_2)$ and
$S(\epsilon_2, \epsilon_1)$ synchronize simultaneously
when the values of $\epsilon_1$ and $\epsilon_2$ are
varied.

In fact, much more is true. Let us define an operator
\begin{eqnarray}\label{eq:identity}
S_t := S({{\epsilon_1+t\epsilon_2}\over {1+t}},
{{t\epsilon_1+\epsilon_2}\over {t+1}})
&=&{{S(\epsilon_1, \epsilon_2)+t S(\epsilon_2, \epsilon_1)}\over {1+t}}.
\end{eqnarray}
In particular $S_0 = S(\epsilon_1, \epsilon_2)$ and
$S_\infty = S(\epsilon_2, \epsilon_1)$. 
Now it can be seen that $S_tX$ lies on a straight line between
$S_0X$ and $S_\infty X$ and all the three points lie on the
same side of the line $x=y$. 
This in particular implies that,
for given values of $\epsilon_1$ and $\epsilon_2$, the 
boundary of the support of the invariant measure of $S_t$
lies between that of $S_0X$ and $S_\infty X$.
In fact, if we
define $D\left(\begin{array}{c}x \\ y
\end{array}\right) = |x-y|$ then we see immediately
that $D(S_t\left(\begin{array}{c}x \\ y
\end{array}\right) ) = |1-\epsilon_1-\epsilon_2||f(x)-f(y)|$
is independent of $t$ and that, consequently,   as $\epsilon_1$ and $\epsilon_2$ are
varied $S_t$ synchronizes independently of $t$. 

Now when $t=1$ we have $S_1 =
S({{\epsilon_1+\epsilon_2}\over 2},{{\epsilon_1+\epsilon_2}\over 2})$
which is symmetric. So we can use the result of the previous section which
tells us that the system synchronizes when $\epsilon_1+\epsilon_2 = 1/2$
and desynchronizes when $\epsilon_1+\epsilon_2 = 3/2$.

It should be noted that the invariant measures generated by 
$S_t$s for different $t$
are not the same. But since we are interested in the 
synchronization property, i.e., the value of $\epsilon$ 
at which the support of the measure shrinks to the line $x=y$,
we use the identity~(\ref{eq:identity}) which gives another
equivalent symmetric dynamics. And since
this equivalent dynamics is symmetric we know its synchronization
properties. 

\section{Globally coupled map network}
Now we extend our analysis to a higher dimensional case.  We consider
a globally coupled network of $N$ maps with all couplings of equal
strength. As a result
\begin{eqnarray}
A=\left( \begin{array}{cccc}
1-\epsilon & \epsilon/(N-1) & \cdots &\epsilon/(N-1) \\
\epsilon/(N-1) &1-\epsilon  & \cdots &\epsilon/(N-1) \\
\vdots & \vdots &&\vdots \\
\epsilon/(N-1) &\epsilon/(N-1) & \cdots &1- \epsilon
\end{array} \right)
\end{eqnarray}
Now we consider the dynamics in the 2-dim space spanned by
$\hat{\bf e}_1$ and $(1,1,\cdots,1)$. By symmetry this dynamics
is the same as that in any other similar subspace, i.e., spanned
by $\hat{\bf e}_i$ and $(1,1,\cdots,1)$. And all of them 
synchronize simultaneously. We take the orthogonal vectors
$\hat{\bf e}_1$ and $(0,1,\cdots,1)$ and a general vector
in this space $X=(x,y,y,\cdots,y)^T$. Now
\begin{eqnarray}
AX=\left( \begin{array}{c}
(1-\epsilon)x + \epsilon y \\
{\epsilon\over{N-1}}x +(1-{\epsilon\over{N-1}})y\\
\vdots \\
{\epsilon\over{N-1}}x +(1-{\epsilon\over{N-1}})y
\end{array} \right).
\end{eqnarray}
As a result, the reduced dynamics of our system in this 2-dim
subspace is given by an effective coupling matrix
\begin{eqnarray}
A_{eff}=\left( \begin{array}{cc}
1-\epsilon & \epsilon \\
{\epsilon\over{N-1}}& (1-{\epsilon\over{N-1}})
\end{array} \right)
\end{eqnarray}
and from the previous section we know that this dynamics
synchronizes when $\epsilon + {\epsilon/(N-1)} = 1/2$,
i.e., $\epsilon = {{N-1}\over{2N}}$. This result again 
agrees with the one obtained by the linear stability 
analysis~\cite{Jalan2003}.

\section{Concluding discussion}
In this paper we have proposed a  method 
that uses invariant measures and in particular
their support to study synchronization. We have
demonstrated it on one example, namely the tent map where 
we were able to study the stability of the support
of the invariant measure using purely geometric arguments.
In the future, different techniques to deal with invariant
measures and their support will have to be developed in
order for this method to be of wide use.

We should note that the notion of convergence we have been studying
here, namely convergence of the support of the invariant measure as
encoded in the convergence of its boundary in general is different
from the weak-$\star$ convergence introduced above. That is, there exists the possibility
that the measure can become zero asymptotically everywhere
except on the synchronization manifold without making the
boundary unstable. In other words, the measure can become
singular. Here, we have considered the 
tent map which is expanding everywhere. But it is not
sufficient to have the individual map expanding, rather
it should be the coupled map that should guarantee the
existence of the absolutely continuous measure. In fact,
it turns out that, in all the cases considered here, the
combined map ceases to be expanding exactly at the
synchronization threshold. 
Clearly this is a consequence of the everywhere expanding
property of the map chosen.
While this guaranteed for us
the existence of the absolutely continuous measure for
$\epsilon < \epsilon_c$ and validated our use of the present
method, it also raises a question as to what happens
in the case of other maps. In fact, 
there exists a considerable activity in Ergodic Theory 
concerning  the existence of an absolutely continuous
invariant measures for various discrete dynamical systems
(see for example~\cite{Gora1989,Arbieto2004,Alves2004}).
So for general mappings our approach will have to be 
complemented by the results in this field.

Other than considering different maps
one can also try to apply this method to more complicated
connection topologies. Various networks, such as, random,
scale-free, small world etc. can be considered.
A characterization of complete
measures is also needed. We have already taken a step
in this direction~\cite{Jost2004b}.
It is important to note that the knowledge of the complete
measure was not necessary to study the synchronization
since we used only the support of the measure. However,
a method to find out that support is the minimum 
prerequisite for this approach to be useful.

\begin{acknowledgments}
One of us (KMK) would like to thank the Alexander-von-Humboldt-Stiftung for financial
support.
\end{acknowledgments}

\bibliography{kolwankar}

\end{document}